\documentclass[aps,pra,twocolumn,superscriptaddress]{revtex4-2}
\usepackage{bbm}
\usepackage{mathrsfs}
\usepackage{amsmath}
\usepackage{amsfonts}
\usepackage[colorlinks=true,linkcolor=red,urlcolor=blue,citecolor=blue,anchorcolor=blue]{hyperref}
\usepackage{graphicx,epstopdf}
\usepackage{subfigure}
\usepackage{epsfig}
\usepackage{dcolumn}
\usepackage{bm}
\usepackage{color}
\usepackage{natbib}
\usepackage{amssymb}
\usepackage{xcolor}
\usepackage{braket}
\usepackage{ulem}
\usepackage{float}
\usepackage{lipsum}

\definecolor{RED}{rgb}{1,0,0}
\definecolor{BLUE}{rgb}{0,0,1}

\definecolor{newtxtcolor1}{rgb}{0, 0, 0}
\newcommand {\rd}[1] {{\color{newtxtcolor1}{#1}}}
\definecolor{movetxtcolor}{rgb}{0, 0, 1}

\definecolor{orangetxtcolor}{rgb}{1, 0.45, 0}

\begin{document}
\title{Metal-Insulator Coexistence and { Gap-Crossing Domain-Wall Modes} in an Aubry-Andr\'{e} Model with Nonlocal Hopping}

\author{Xiarui Zhan}
\thanks{These authors contributed equally to this work.}
\address{State Key Laboratory of Artificial Microstructure and Mesoscopic Physics, School of Physics, Frontiers Science Center for Nano-optoelectronics, $\&$ Collaborative Innovation Center of Quantum Matter, Peking University, Beijing 100871, China}

\author{Mingsheng Tian}
\thanks{These authors contributed equally to this work.}
\address{Department of Physics, The Pennsylvania State University, University Park, Pennsylvania, 16802, USA}
\address{State Key Laboratory of Artificial Microstructure and Mesoscopic Physics, School of Physics, Frontiers Science Center for Nano-optoelectronics, $\&$ Collaborative Innovation Center of Quantum Matter, Peking University, Beijing 100871, China}

\author{Qiongyi He}
\address{State Key Laboratory of Artificial Microstructure and Mesoscopic Physics, School of Physics, Frontiers Science Center for Nano-optoelectronics, $\&$ Collaborative Innovation Center of Quantum Matter, Peking University, Beijing 100871, China}
\address{Hefei National Laboratory, Hefei, 230088, China}
\address{Collaborative Innovation Center of Extreme Optics, Shanxi University, Taiyuan 030006, China}

\author{Kaiye Shi}
\email{sky@pku.edu.cn}
\address{State Key Laboratory of Artificial Microstructure and Mesoscopic Physics, School of Physics, Frontiers Science Center for Nano-optoelectronics, $\&$ Collaborative Innovation Center of Quantum Matter, Peking University, Beijing 100871, China}
\address{Hefei National Laboratory, Hefei, 230088, China}

\author{Wei Zhang}
\email{wzhangl@ruc.edu.cn}
\affiliation{Department of Physics, Renmin University of China, Beijing 100872, China}
\affiliation{Beijing Academy of Quantum Information Sciences, Beijing 100193, China}

\begin{abstract}
Nonequilibrium transport remains a central theme in modern physics, spanning from condensed matter to synthetic systems. Here, we investigate particle transport in an extended Aubry-André model with { system-scale hopping, namely nonlocal hopping with a range proportional to the system size}, and uncover \rd{a metal-insulator coexistence regime in real space, where metallic and insulating spatial domains coexist within the same system and are separated by sharp spatial boundaries}. In the insulating region, particles exhibit flat-band-like localization in the absence of quasiperiodic potentials, while a quasiperiodic potential induces distinct multi-point localization, \rd{different from} conventional exponential localization. Meanwhile, particles can freely propagate and tunnel across \rd{spatially disconnected metallic domains} separated by the insulating region. Beyond this coexistence phase, we identify { unconventional gap-crossing domain-wall modes} with comb-like spatial profiles that mediate nonlocal, multi-point transport across separated metallic domains. Our findings reveal a rich interplay between { localization, nonlocality, and { transport}} in systems with nonlocal hopping.
\end{abstract}

\maketitle
\section{Introduction}
\label{sec:intro}

Over half a century ago, Anderson demonstrated that random disorder can lead to the exponential localization of single-particle wavefunctions, a fundamental quantum phenomenon now known as Anderson localization~\cite{Anderson1958}, which has since aroused intensive discussion~\cite{Lee1985}. Building on this, scaling theory~\cite{Abrahams1979} later revealed that in one and two dimensions, arbitrarily weak disorder suffices to localize all eigenstates, while in three dimensions, extended and localized states may coexist and are separated by a critical energy known as the mobility edge~\cite{Evers2008}. This enables a novel disorder-driven metal-insulator transition when the Fermi energy crosses the mobility edge. Beyond random disorder, quasiperiodic potentials can also induce localization and support mobility edges even in one-dimensional systems~\cite{Das1988,Biddle2010}, generating growing theoretical and experimental interest~\cite{Roati2008,Luschen2018,Yao2019}. 
{ A paradigmatic example is the Aubry-Andr\'{e} (AA) model~\cite{Harper1955,Aubry1980}. In its standard nearest-neighbor form, self-duality produces a global transition between fully extended and fully localized single-particle states, without an energy-dependent mobility edge.
AA-type localization has been observed in ultracold-atom optical lattices and photonic lattices~\cite{Roati2008,Lahini2009}. Beyond the standard nearest-neighbor AA model, deviations from the deep tight-binding limit can generate energy-dependent localization transitions, enabling the experimental observation of single-particle mobility edges in bichromatic optical lattices~\cite{Luschen2018}. Theoretically, mobility edges have been realized in various generalized AAH models by introducing exponentially decaying non-nearest-neighbor hopping~\cite{Biddle2010}, deforming the quasiperiodic potential while retaining a generalized self-duality~\cite{Ganeshan2015}, considering shallow quasiperiodic continuum lattices~\cite{Yao2019}, or applying quasiperiodic modulations only to selected lattice sites in mosaic models~\cite{Wang2020}.
Beyond single-particle localization, quasiperiodic AA-type systems have also become important platforms for topological pumping~\cite{Kraus2012,Verbin2013}.
}

Previous works on quasiperiodic models mostly focus on local systems with nearest-neighbor hopping~\cite{Wang2020,Zhou2023}. Recent advances in quantum simulation platforms, such as Rydberg atom arrays~\cite{Saffman2010}, trapped ions~\cite{Monroe2021}, and polar molecules~\cite{Yan2013}, have brought increasing attention to systems with long-range coupling~\cite{Defenu2023}. In particular, long-range extensions of the AA model, featuring power-law decayed hopping, have been shown to host an ergodic-to-multifractal transition~\cite{Deng2019}. 
Recently, both theoretical and experimental efforts have explored the more exotic extreme limit of long-range coupling, i.e., nonlocal interaction (or hopping) of the same scale as the system. { For the hopping model studied below, we refer to this regime as system-scale hopping: the hopping range $R$ scales linearly with the system size $N$, i.e., $R/N=O(1)$.} This form of coupling has been shown to enable a host of intriguing phenomena, including enhanced metrological sensitivity~\cite{norcia2018,luo2025,luo2024,perlin2020,block2024-metrology,Chen2024,chu2023-metrology}, exotic quantum phases and anomalous dynamics~\cite{landig2016-phase,mottl2012-phase,tian2023,sky2023nbt,Qiao2024}, and singular entanglement behavior~\cite{sharma2022-ent}.
While system-scale hopping is difficult to realize in natural materials, artificial platforms, such as electrical circuits~\cite{Lu2018,wang2023,albert2015a-circuit} and synthetic metamaterials~\cite{chen2025a-bryce,meier2018-bryce,Lustig2019,Tian2024,anandwade2023-mechanical,Wang2024,Tian2025}, offer flexible architectures where almost arbitrary coupling profiles can be implemented and fine tuned. These platforms thus provide a powerful avenue for exploring emergent phenomena in nonlocal systems with system-scale hopping.

In this work, we investigate how system-scale hopping influences quantum dynamics of particles in an extended AA model (Fig.~\ref{fig1}). Such hopping partitions the system into distinct spatial regions, presenting different responses to the quasiperiodic potential. This spatial inhomogeneity gives rise to a well-defined \rd{metal-insulator (MI) coexistence phase}, characterized by the coexistence of insulating \rd{spatial domains} dominated by localized single-particle dynamics and metallic \rd{spatial domains} where the wavefunction remains extended. In the insulating region, the wavefunction displays a form of multi-point localization, \rd{in sharp contrast to} conventional exponential localization. The metallic regions support extended wavefunctions spanning multiple, spatially disconnected domains. Remarkably, the \rd{MI coexistence phase} persists even in the absence of quasiperiodic potential, owing to the presence of a flat band with degenerate eigenstates.
Furthermore, we uncover { anomalous gap-crossing domain-wall modes} uniquely induced by system-scale hopping. These modes exhibit comb-like spatial probability distributions, enabling coherent transmission between spatially separated metallic regions, which are fundamentally distinct from conventional edge-localized modes.

\section{Metal-insulator coexistence phase}

\begin{figure}[tbp]
	\centering
	\includegraphics[width=1\linewidth]{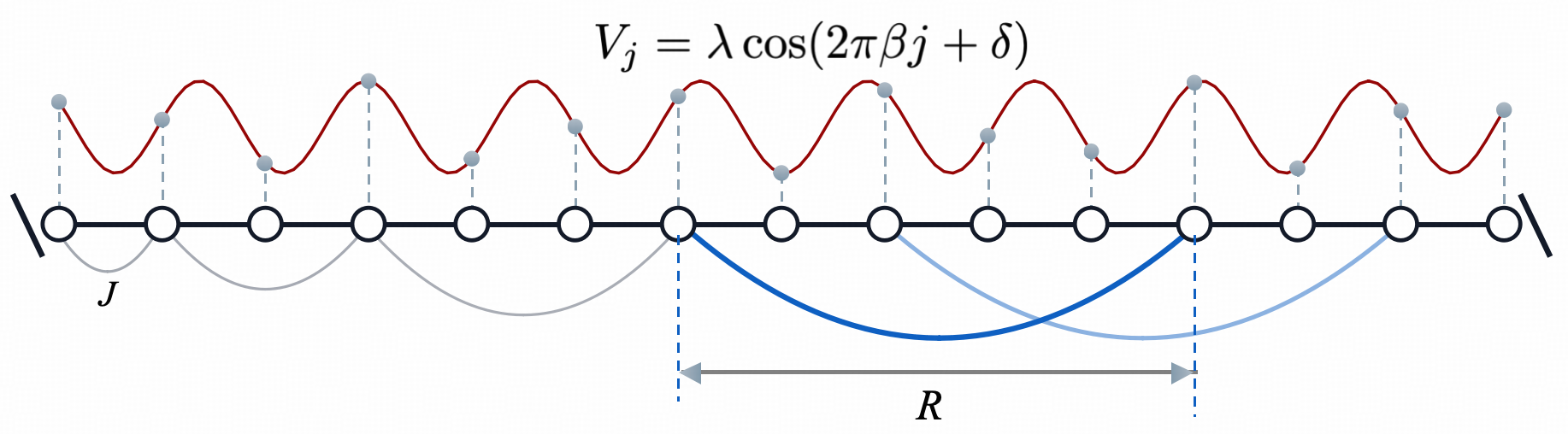}
	\caption{\rd{Schematic of the extended Aubry-Andr\'{e} model in Eq.~(\ref{Hamiltonian}). Open circles denote the sites of an open one-dimensional chain. The red curve and dashed vertical lines indicate the quasiperiodic onsite potential $V_j=\lambda\cos(2\pi\beta j+\delta)$. The arcs highlight hopping processes of amplitude $J$ between sites separated by distances up to the hopping range $R$.}}
	\label{fig1} 
\end{figure}

We consider an extended AA model (Fig.~\ref{fig1}) with Hamiltonian
\begin{eqnarray}
\label{Hamiltonian}
H &=&  J\sum_{l=1}^{R} \sum_{j} (a^{\dagger}_ja_{j+l}+\mathrm{H.c.} )
+\sum_j V_ja^{\dagger}_ja_{j},
\end{eqnarray}
where $a_j$ and $a^{\dagger}_j$ represent particle operators of the $j$-th site, $J$ denotes the hopping amplitude with range $R$, and the quasiperiodic potential $V_j=\lambda\cos{(2\pi \beta j +\delta)}$ is characterized by strength $\lambda$, phase shift $\delta$ and an irrational number $\beta={(\sqrt{5}-1)}/{2}$. 
\rd{The box-like hopping profile is used here as a minimal model of the extreme nonlocal regime with $R/N=O(1)$. It provides an analytically transparent way to isolate the effect of boundary-induced connectivity inhomogeneity, and can be implemented in synthetic platforms where coupling matrices are programmable~\cite{Wang2024,anandwade2023-mechanical,Chen2024,chen2025a-bryce}.
Unless stated otherwise, all discussions that follow assume open boundary conditions. This choice is essential for the MI coexistence physics discussed below: with periodic boundary conditions every \rd{site} has the same connectivity for the constant hopping range $R$, so the boundary-induced spatial partition disappears.}

Assuming the particle starts at site $x_0$, the time-evolved quantum state can be expressed as $\ket{\psi,t} = \sum_{n} e^{-iE_n t}\ket{\phi_n} \langle \phi_n|x_0\rangle  $, where $\ket{x_0}=a^{\dagger}_{x_0}\ket{\mathrm{vac}}$ with vacuum state $\ket{\mathrm{vac}}$, and $\ket{\phi_n}$ and $E_n$ are the single-particle eigenstates and eigenenergies of the Hamiltonian (\ref{Hamiltonian}), respectively, satisfying $H\ket{\phi_n}=E_n\ket{\phi_n}$.
{ 
Before turning to the extreme nonlocal regime, we recall the standard AA model ($R=1$). In this case, the Hamiltonian (\ref{Hamiltonian}) is self-dual at $\lambda=2J$ via the transformation $\ket{j} = \frac{1}{\sqrt{N}}\sum_{\bar{k}}e^{i(2\pi\beta j + \delta) \bar{k}}e^{i\bar{\delta}j}\ket{\bar{k}}$~\cite{Aubry1980}}.
Since this transformation maps extended (localized) states in real space to localized (extended) states in momentum space, the self-dual point marks the transition from extended to localized states and is referred to as the AA self-duality point. 
When $\lambda <2J$ ($\lambda >2J$), all eigenstates are extended (localized), and the system is in the metallic (insulating) phase, where particles can propagate throughout the whole system (remain localized in space). { For a normalized eigenstate $\ket{\phi_n}=\sum_j u^{(n)}_{j}a^{\dagger}_j\ket{\mathrm{vac}}$, where $n$ labels the eigenstate, its localization character can be diagnosed by the inverse-participation-ratio-based fractal dimension $D_n$~\cite{Evers2008,Wang2020,Zhou2023}, defined as
\begin{equation}
D_n=-\lim_{N\rightarrow\infty}\frac{\ln(\mathrm{IPR}_n)}{\ln N},
\qquad
\mathrm{IPR}_n=\sum_j|u_j^{(n)}|^4 .
\end{equation}
The participation ratio $\mathrm{PR}_n=1/\mathrm{IPR}_n$ estimates the effective participation volume of the wavefunction. If its weight is distributed approximately uniformly over $M_n$ sites, then $\mathrm{IPR}_n\sim M_n^{-1}$, so that $M_n\sim N^{D_n}$. In one dimension, an extended state has $D_n=1$, whereas an exponentially localized state has $D_n=0$. 

}

\begin{figure}[tbp]
	\centering
	\includegraphics[width=1\linewidth]{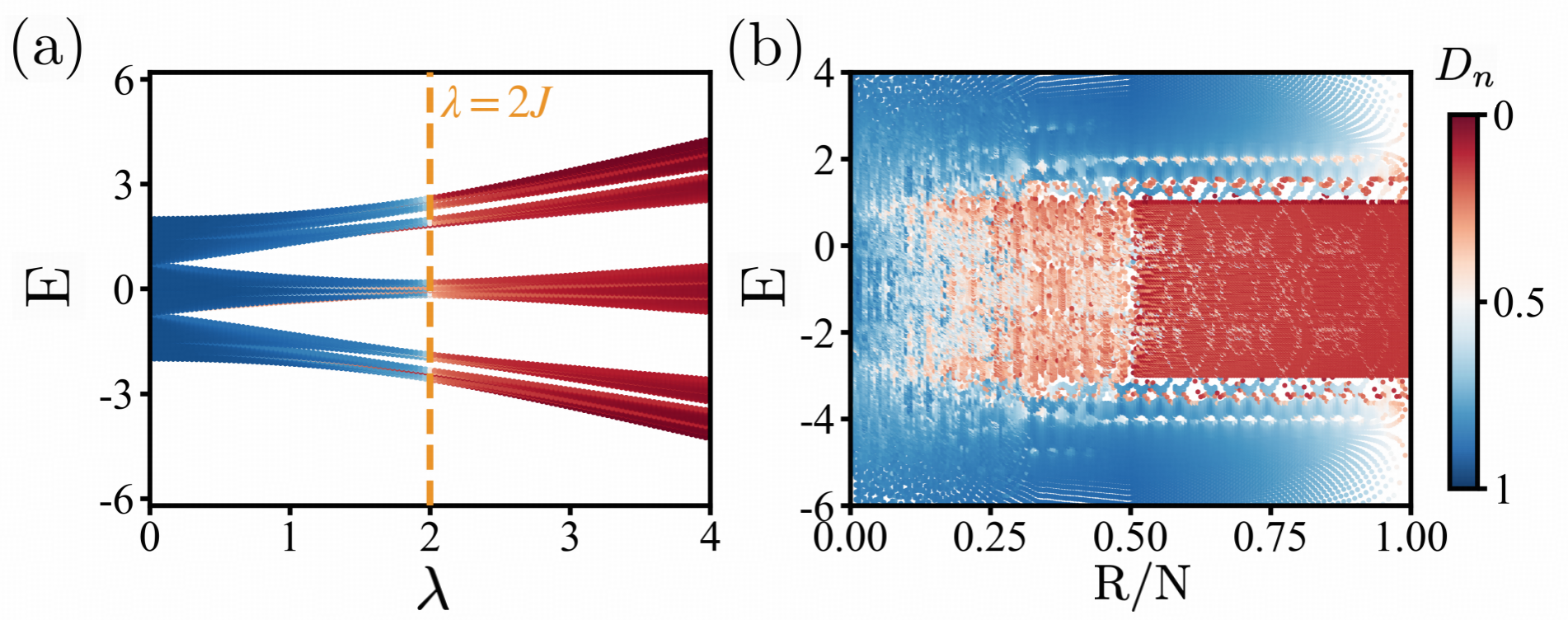}
	\caption{{ Fractal dimension around the conventional Aubry-Andr\'{e} self-duality point.
(a) Spectrum of the nearest-neighbor AA model ($R=1$) as a function of $\lambda/J$ for $N=1583$; the orange dashed line marks $\lambda=2J$.
(b) Spectrum as a function of $R/N$ at $\lambda=2J$ for $N=1584$.
The color denotes the fractal dimension $D_n$. The energy unit is set to $J=1$.}}
	\label{fig2}
\end{figure}

{ Figure~\ref{fig2}(a) shows the nearest-neighbor AA limit ($R=1$), where the spectrum changes from extended states for $\lambda<2J$ to localized states for $\lambda>2J$, with the self-dual point at $\lambda=2J$. Figure~\ref{fig2}(b) fixes $\lambda=2J$ and varies the hopping range $R$. As $R/N$ increases, the spectrum no longer remains a globally critical AA spectrum. Instead, states with small $D_n$ appear in a finite energy window and form a localized sector, while states with large $D_n$ remain in the surrounding spectral regions. This result shows that, in the large-$R$ regime, the system no longer exhibits a single global localization phase. Instead, energy boundaries separating localized and extended states emerge, namely mobility edges, as discussed in more detail in the next subsection.}
\subsection{Partitioned dynamics}

\begin{figure}[tbp]
	\centering
	\includegraphics[width=1\linewidth]{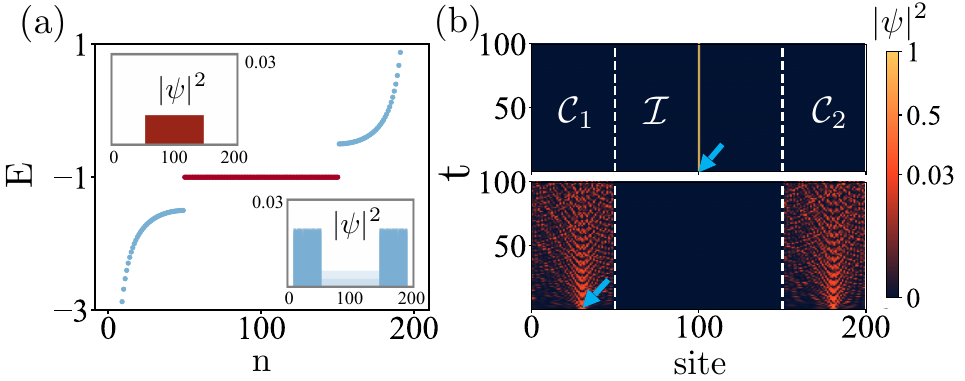}
	\caption{
   Single-particle energy spectra and their dynamical behavior in the presence of system-scale hopping.  
(a) A three-branch spectral structure featuring a red intermediate band and blue side bands. The insets in the upper left and bottom right corners display the wavefunction distributions of the intermediate band and side bands, respectively. 
(b) Time evolution of the particle density distribution for initial positions at $x_0 = 100$ \rd{(site region ${\cal I}$)} and $x_0 = 30$ \rd{(site region ${\cal C}_1$)}. Blue arrows indicate the initial positions.
    Here, { we set $J=1$; the other parameters are} $ \lambda=0$, $R=150$, and $N=200$.}
	\label{fig3} 
\end{figure}

Next, we consider \rd{system-scale hopping with $R/N=O(1)$, or equivalently $R=\alpha N$ with $0<\alpha<1$}. \rd{We first discuss} the case without incommensurate potential ($\lambda =0$). 
The single-particle spectrum exhibits a three-branch structure [Fig.~\ref{fig3}(a)] with two ``side bands'' (blue) and an intermediate ``flat band'' (red). All eigenstates of the intermediate band are confined within \rd{the site region} ${\cal I}=(N-R,R)$ for $R>N/2$ [upper inset of Fig.~\ref{fig3}(a)], and are referred to as intermediate-band modes $\ket{\phi^{\mathrm{I}}_{n_{\mathrm{I}}}}$ \rd{(IB modes)}. 
In contrast, side bands contain a large number of eigenstates confined in \rd{the site region} ${\cal C} =(1,N-R)\cup (R,N)$ [lower inset of Fig.~\ref{fig3}(a)], denoted as clustered side-band modes $\ket{\phi_{n_{\mathrm{C}}}^{\mathrm{C}}}$ \rd{(CSB modes)}, as well as some trivial side-band modes \rd{(TSB modes)} $\ket{\phi^{\mathrm{T}}_{n_{\mathrm{T}}}}$ distributed throughout the entire chain. 
We denote the numbers of \rd{IB modes, CSB modes, and TSB modes} as $N_{\mathrm{I}}$, $N_{\mathrm{C}}$, and $N_{\mathrm{T}}$, respectively.
For example, when $R>N/2$, we have $N_{\mathrm{T}} \ll N_{\mathrm{I}},N_{\mathrm{C}}$, and the evolution of an initial state $| x_0 \rangle$ is dominated by the \rd{IB and CSB modes}, with $\ket{\psi,t} \approx \sum_{n_{\mathrm{I}}}e^{-iE_{n_{\mathrm{I}}}t} \ket{\phi^{\mathrm{I}}_{n_{\mathrm{I}}}} \langle \phi^{\mathrm{I}}_{n_{\mathrm{I}}} | x_0\rangle+ \sum_{n_{\mathrm{C}}}e^{-iE_{n_{\mathrm{C}}}t} \ket{\phi^{\mathrm{C}}_{n_{\mathrm{C}}}} \langle \phi^{\mathrm{C}}_{n_{\mathrm{C}}} | x_0\rangle$. 

If $x_0$ is located in \rd{the site region} ${\cal I}$ [upper panel of Fig.~\ref{fig3}(b)], $\ket{\psi,t} \approx \sum_{n_{\mathrm{I}}} e^{-iE_{n_{\mathrm{I}}} t}\ket{\phi^{\mathrm{I}}_{n_{\mathrm{I}}}} \langle \phi^{\mathrm{I}}_{n_{\mathrm{I}}}|x_0\rangle  \approx e^{iJt}\ket{x_0}$.
This means that the flat band prohibits transport, causing the particle to remain almost completely localized at the initial position. This localization mechanism, later referred to as flat-band-like localization, differs from disorder-induced localization and arises from the interplay between the intermediate flat spectral structure and the spatially inhomogeneous wavefunction distribution. On the other hand, when the initial position of the particle is in \rd{the site region} ${\cal C}$, the evolution dynamics is dominated by \rd{CSB modes}, with $\ket{\psi,t} \approx \sum_{n_{\mathrm{C}}}e^{-iE_{n_{\mathrm{C}}}t} \ket{\phi^{\mathrm{C}}_{n_{\mathrm{C}}}} \langle \phi^{\mathrm{C}}_{n_{\mathrm{C}}} | x_0\rangle$.
The \rd{CSB modes} exhibit a relatively uniform density distribution [lower inset of Fig.~\ref{fig3}(a)], being spatially separated into two parts ${\cal C}_1=(1,N-R)$ and ${\cal C}_2=(R,N)$ by \rd{the site region} ${\cal I}$. The nonlocal correlation between ${\cal C}_1$ and ${\cal C}_2$ allows coherent tunneling, i.e., even if the initial position is confined in ${\cal C}_1$, the particle still has a high probability to rapidly appear in ${\cal C}_2$. As demonstrated in the lower panel of Fig.~\ref{fig3}(b), the density distribution of a particle starting from ${\cal C}_1$ not only spreads in ${\cal C}_1$, but also evolves in ${\cal C}_2$ in a similar manner, with negligible distribution in \rd{the site region} ${\cal I}$. 

\begin{figure}[tbp]
	\centering
	\includegraphics[width=1\linewidth]{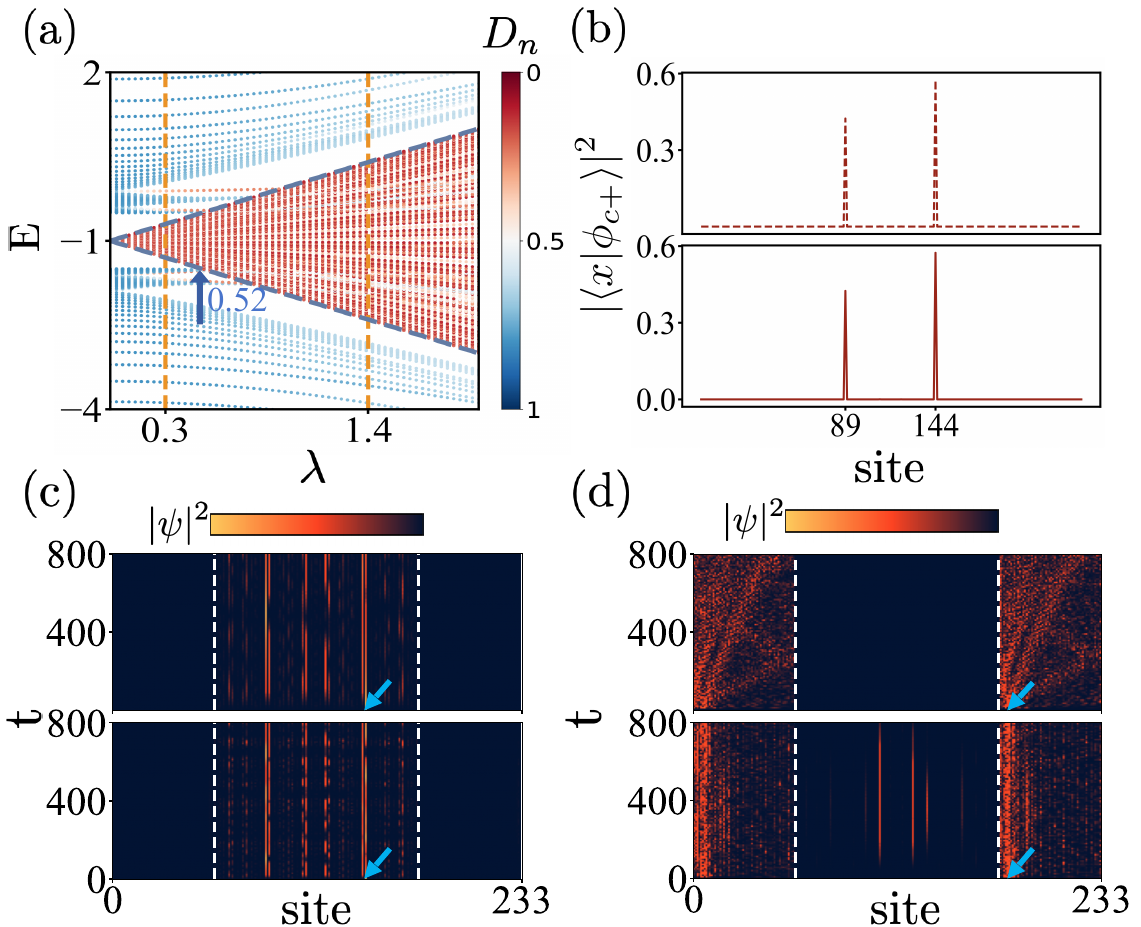}
	\caption{Energy spectra, localized state and dynamical behaviors in the presence of quasiperiodic potential.
(a)~The energy spectra and fractal dimension as a function of $\lambda$ \rd{starting from $\lambda/J=0$}, where the color indicates the values of the fractal dimensions $D_n$.
The width of the intermediate band increases linearly with $\lambda$, and the boundary (blue dashed line) separates the extended and localized states, forming mobility edges with $E_{c\pm} = \pm\lambda - J$.
The corresponding numerical localized state $\ket{\phi_{c+}}$ with $E_n=E_{c+}$ \rd{at the upper boundary} is shown in the upper panel of (b), and agrees well with the analytical solution (lower), where the localized states are shown to be related to the Fibonacci numbers ($89$, $144$), with the system size being $N = 233$.
(c)~The time evolution of the density distribution of the particle initially excited in \rd{the site region} ${\cal I}$ ($x_0 = 144$) under weak (upper) and strong (lower) quasiperiodic potentials with $\lambda = 0.3$ and $\lambda = 1.4$, respectively.
(d)~Same conditions as in (c), but with the particle initially excited in \rd{the site region} ${\cal C}$ ($x_0 = 180$). Here, we take $R=144$, $\delta=0$, and \rd{$J=1$} in all plots.
}
	\label{fig4}
\end{figure}

The different transport properties of particles in different \rd{site regions indicate an MI coexistence phase}. This phase hosts \rd{sharp spatial boundaries} [white lines of Fig.~\ref{fig3}(b)], separating the metallic \rd{site region} ${\cal C}$ (particles can move freely) and the insulating \rd{site region} ${\cal I}$ (particle transport is suppressed). 
This disorder-free \rd{MI coexistence phase arises from the joint action of the spatially inhomogeneous eigenstate distribution} and flat spectral structure. 
\rd{It relies on open boundary conditions: sites near the physical boundaries have fewer hopping partners than sites in the central part of the chain, which produces distinct connectivity domains for $R/N=O(1)$. Under periodic boundary conditions, this connectivity inhomogeneity is removed and the corresponding MI coexistence phase disappears.} Given that free-particle localization typically arises from disorder or quasiperiodic potentials, a more general \rd{MI coexistence phase} is expected to exist in systems with disorder (or quasiperiodic potentials) and system-scale hopping, as discussed below.

\subsection{Multi-point localized state}

We now turn to the case with quasiperiodic potential ($\lambda >0$). Figure~\ref{fig4}(a) shows how the eigenvalues and their associated fractal dimensions $D_n$ evolve. As $\lambda$ increases, the \rd{IB modes} are localized and the spectral width grows linearly, forming a conical structure. The boundary of the \rd{IB-mode} cone marks the mobility edges, located at $E_{c\pm} = \pm \lambda - J$.
States between the mobility edges ($E_{c-} < E_n < E_{c+}$) are localized, while those outside ($E_n < E_{c-}$ or $E_n > E_{c+}$) are extended.

In the regime of weak quasiperiodic potential $\lambda < 0.52$ [Fig.~\ref{fig4}(a)], the energy gaps between the intermediate and side bands protect the spatial separation between the \rd{IB and CSB modes}. As a result, the \rd{IB modes} remain confined within \rd{the site region} ${\cal I}$. { This spatial confinement allows us to project the eigenvalue equation onto ${\cal I}$ and obtain the analytical form of the localized wavefunction.
We write an intermediate-band eigenstate as
\begin{equation}
|\phi_{n}^{I}\rangle=\sum_{j=N-R}^{R} c_{nj} |j\rangle,
\end{equation}
where \(|j\rangle =a_j^{\dagger}\ket{\mathrm{vac}}\) and $c_{nj}=\langle j|\phi_{n}^{I}\rangle$. Substituting this form into the eigenvalue equation and comparing the coefficients of $\ket{j}$ with $j\in {\cal I}$, we obtain
\begin{equation}
\sum_{k=N-R}^{R} c_{nk} = \left[E_n/J + 1 - \lambda/J \cos(2\pi \beta j+\delta)\right] c_{nj}.
\label{eq:coeff_compare_main}
\end{equation}
The left-hand side, $\sum_{k=N-R}^{R} c_{nk}$, is independent of the site index. Therefore, all the site dependence of $c_{nj}$ is contained in the denominator, giving
\begin{equation}
	c_{nj} \propto \frac{1}{E_n + J - \lambda \cos(2\pi \beta j +\delta)}. 
\end{equation}
With that, we obtain an analytical expression of the wavefunction,
\begin{equation}
\label{psi}
|\phi_{n}^{I}\rangle=\frac{1}{{\cal N}_I}{\sum_{j=N-R}^{R}} \frac{\ket{j}}{E_n + J - \lambda \cos(2\pi \beta j +\delta)},
\end{equation}
with the normalization constant ${\cal N}_I$.}

Unlike Anderson localization, Eq.~(\ref{psi}) indicates that the wavefunction exhibits a multi-peak spatial structure. 
{ Peaks occur at sites minimizing the absolute value of denominator in Eq.~(\ref{psi}), i.e., satisfying $E_n+J\simeq\lambda\cos(2\pi\beta j+\delta)$. As an example, Fig.~\ref{fig4}(b) (with $\delta=0$ throughout) displays the wavefunction for $E_n=\lambda-J$, for which the peak condition becomes $\cos(2\pi\beta j)\simeq 1$.}
Choosing the system size $N$ as the Fibonacci number $F_m = 233$ (with $m = 13$, defined by $F_0 = 0$, $F_1 = 1$, and $F_m = F_{m-1} + F_{m-2}$), the wavefunction is peaked at $j_l = F_{m-l}$, for integers $l$ satisfying $ F_{m-l} \in {\cal I}$. In this case, $j_1 = F_{12} = 144$ and $j_2 = F_{11} = 89$. The amplitudes satisfy $|\psi_{144}| > |\psi_{89}|$ since \rd{$F_{n-1}/F_n \to \beta$} as $n$ increases.
For more general cases with $E_n \neq \lambda - J$, more peaks appear in the wavefunction of \rd{IB modes}. We refer to such states as multi-point localized states, which lead to the distinct propagation structures observed in the single-particle dynamics within \rd{the site region} ${\cal I}$ [Fig.~\ref{fig4}(c)].

On the other hand, the properties of \rd{CSB modes} remain largely unchanged for small $\lambda$, and the dynamics in \rd{the site region} ${\cal C}$ still exhibit extended properties as shown in the upper panel of Fig.~\ref{fig4}(d). However, as $\lambda$ increases, some extended states gradually become localized and eventually cross the mobility edges to enter the localized region [Fig.~\ref{fig4}(a)]. The presence of such states leads \rd{to} a mixed behavior of localization and diffusion for particles placed within ${\cal C}$.
The lower panel of Fig.~\ref{fig4}(d) shows the time-evolved spatial distribution of a particle initially placed in \rd{the site region} ${\cal C}_2$. First, owing to the presence of localized \rd{CSB modes}, the particle exhibits a high probability near its initial site $x_0$ in ${\cal C}_2$ and the corresponding site $x_0 - R$ in ${\cal C}_1$. At the same time, it maintains a nonzero probability of diffusing across the entire ${\cal C}$ region due to the extended \rd{CSB modes}. Moreover, since there is no gap between \rd{CSB and IB modes}, states both inside and outside the cone can be excited, such that there exists a finite probability to find the particle inside \rd{the site region} ${\cal I}$, indicating that the \rd{CSB modes} entering the localized zone can \rd{hybridize with} the \rd{IB modes} and disrupt the regional separation. Consequently, the wavefunction is no longer confined to either ${\cal C}$ or ${\cal I}$, but instead spans both regions. This breakdown of spatial segregation signals the transition from \rd{the MI coexistence phase} to a fully localized insulating phase.

\begin{figure}[tbp]
	\centering
	\includegraphics[width=1\linewidth]{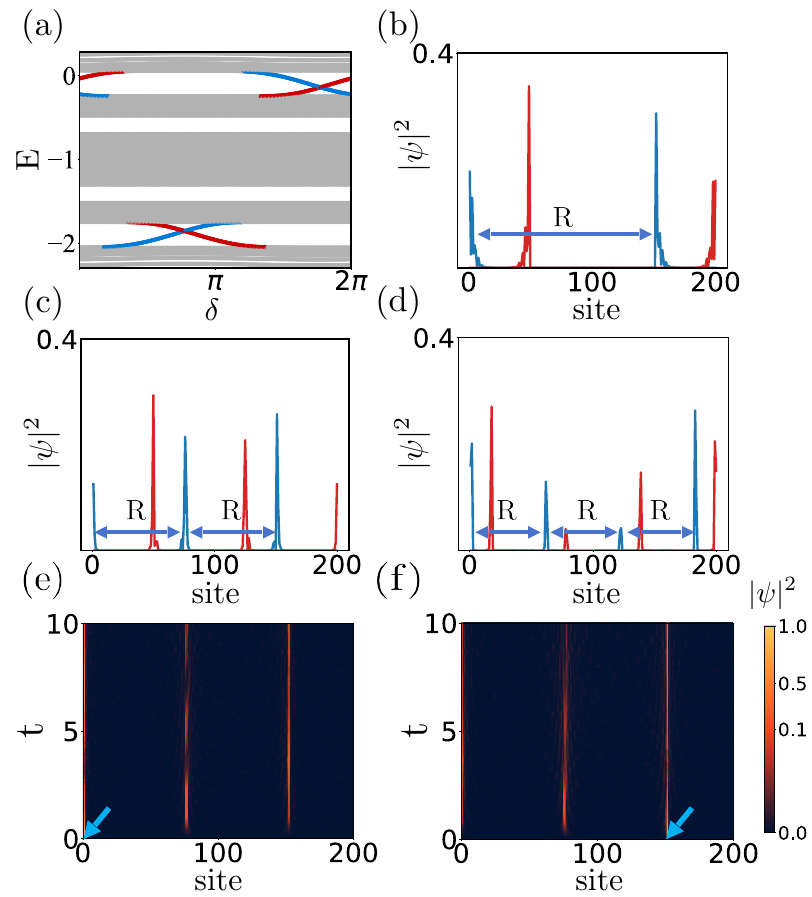}
	\caption{ { Multipartite distributions of gap-crossing domain-wall modes.
    (a) The energy spectra as a function of the phase shift $\delta$.  
  { Clear energy gaps and gap-traversing branches are observed.} The red and blue lines represent different gap-crossing modes, and the corresponding wavefunction distributions are shown in (b), represented by the same color.  The hopping range $R = 150$ for (a) and (b).
  (c) and (d) show real-space distributions of gap-crossing domain-wall modes for $R = 75$ and $R = 60$, featuring $m = 3$ and $m = 4$ equidistant peaks, respectively.
  (e,f) Time-evolved density profiles of gap-crossing domain-wall modes with $m=3$, where the blue arrows denote the initially excited positions.}
  All panels use \rd{$J=1$,} $N=200$ and $\lambda=0.3$; (b)-(f) additionally set phase shift $\delta=0.1$.}
	\label{fig5} 
\end{figure}

\section{\texorpdfstring{{ Gap-crossing domain-wall modes}}{Gap-crossing domain-wall modes}}

{ The AA model is also known to exhibit spectral flow related to topological pumping.} By interpreting the phase shift $\delta$ as a transverse quasimomentum $k_y$, the extended AA model (\ref{Hamiltonian}) can be formally mapped to a two-dimensional Hofstadter-type Hamiltonian~\cite{Hofstadter1976,Kraus2012}, with arbitrary-range hopping along the $x$-direction (AA chain) and periodicity in the $k_y$ direction:
\begin{eqnarray} \label{Hky}
	H(k_y) &=&  J\sum_{l=1}^{R} \sum_{j} (a^{\dagger}_{j,k_y}a_{j+l,k_y}+\mathrm{H.c.} )
	\nonumber \\
	&&+ \sum_j \lambda \cos{(2\pi\beta j+k_y)}a^{\dagger}_{j,k_y}a_{j,k_y}.
\end{eqnarray}
This two-dimensional Hamiltonian $H^{\rm 2D} = \sum_{k_y} H(k_y)$ features { gapped spectra with gap-traversing branches} in the system-scale hopping regime [Fig.~\ref{fig5}(a)].
{ We have tested the conventional filled-subspace Chern-marker diagnostic for these gaps. While the same diagnostic reproduces the expected integer result for the standard nearest-neighbor AA benchmark, it does not yield a stable integer-valued bulk plateau for the system-scale hopping regime. We therefore do not assign a conventional bulk Chern number to these modes, and instead characterize them by their gap-crossing spectral flow and real-space domain-wall structure.}

Remarkably, in contrast to conventional edge modes that are exponentially localized near the system boundaries, the { gap-crossing domain-wall modes} in the { system-scale hopping case} exhibit multipartite distributions [Figs.~\ref{fig5}(b-d)], enabling coherent, nonlocal transport across spatially disconnected metallic regions [Figs.~\ref{fig5}(e,f)]. 
These phenomena can be attributed to the interplay between { gap-crossing spectral flow} and nonlocality introduced by system-scale hopping. 
As shown above, when the hopping range { scales with the system size}, the system naturally partitions into distinct regions. { The gap-crossing modes}, which bridge the gaps, are no longer confined to the ends of the chain. Instead, they can emerge at spatial-domain boundaries $j = mR$ and $j = N - mR$ ($m \in \mathbb{Z}$), and are therefore referred to as { gap-crossing domain-wall modes}.

These states exhibit multiple, spatially periodic peaks that form a comb-like structure in their spatial profile, consisting of $m$ equidistant peaks with spacing $R$ [Figs.~\ref{fig5}(b-d)].
This comb-like structure enables coherent, nonlocal, multi-point quantum transport, i.e., a particle initialized at any one of the peaks can coherently tunnel to the others, resulting in highly nonlocal dynamic correlations. This is evidenced by the time-evolved density profiles shown in Figs.~\ref{fig5}(e,f).
These findings \rd{show} how system-scale hopping can qualitatively reshape the spatial character of { gap-crossing modes}, opening new avenues for engineering { nonlocal gap-crossing transport} in synthetic systems.

{ 
\section{Experimental platforms}
\label{sec:experiment}

Our extended Aubry-Andr\'{e} model can be implemented in several artificial synthetic simulators. Here we discuss experimental realization schemes for two platforms: (A) mechanical oscillators~\cite{Tian2024,anandwade2023-mechanical}, and (B) topoelectrical circuit systems~\cite{Lu2018,wang2023,albert2015a-circuit}.

\subsection{Mechanical oscillators}

The essential idea of using the mechanical system to study the extended AA model is to map the Heisenberg equation generated by a target Hamiltonian onto Newtonian equations of motion for classical oscillators in phase space~\cite{anandwade2023-mechanical,Sirota2020,Tian2024}.
As shown in Fig.~\ref{fig6}, the setup includes $N$ identical mechanical oscillators, whose positions and momenta are continuously measured. By applying individual feedback forces to each oscillator in response to real-time measurements, the couplings between different oscillators can be constructed, allowing us to simulate target Hamiltonians.
A desired Hamiltonian, denoted by $\hat{H}$, is achieved by applying feedback forces
according to the equation $F_i = dp_i/dt = -\partial \hat{H} / \partial x_i$, where $x_i$ denotes the displacement of the $i$-th oscillator.
Naturally, self-feedback forces proportional to the oscillator positions ($F_i \propto x_i$) shift their frequencies by $\Delta \omega$ from a natural frequency. Nonlocal hopping terms are induced by the cross-feedback forces related to the position of oscillators at distant locations ($F_i \propto x_{j}$). The oscillator-specific control of feedback forces allows for the design of structured, nonlocal discrete-site models.

\begin{figure}[tb]
	\centering
	\includegraphics[width=1\linewidth]{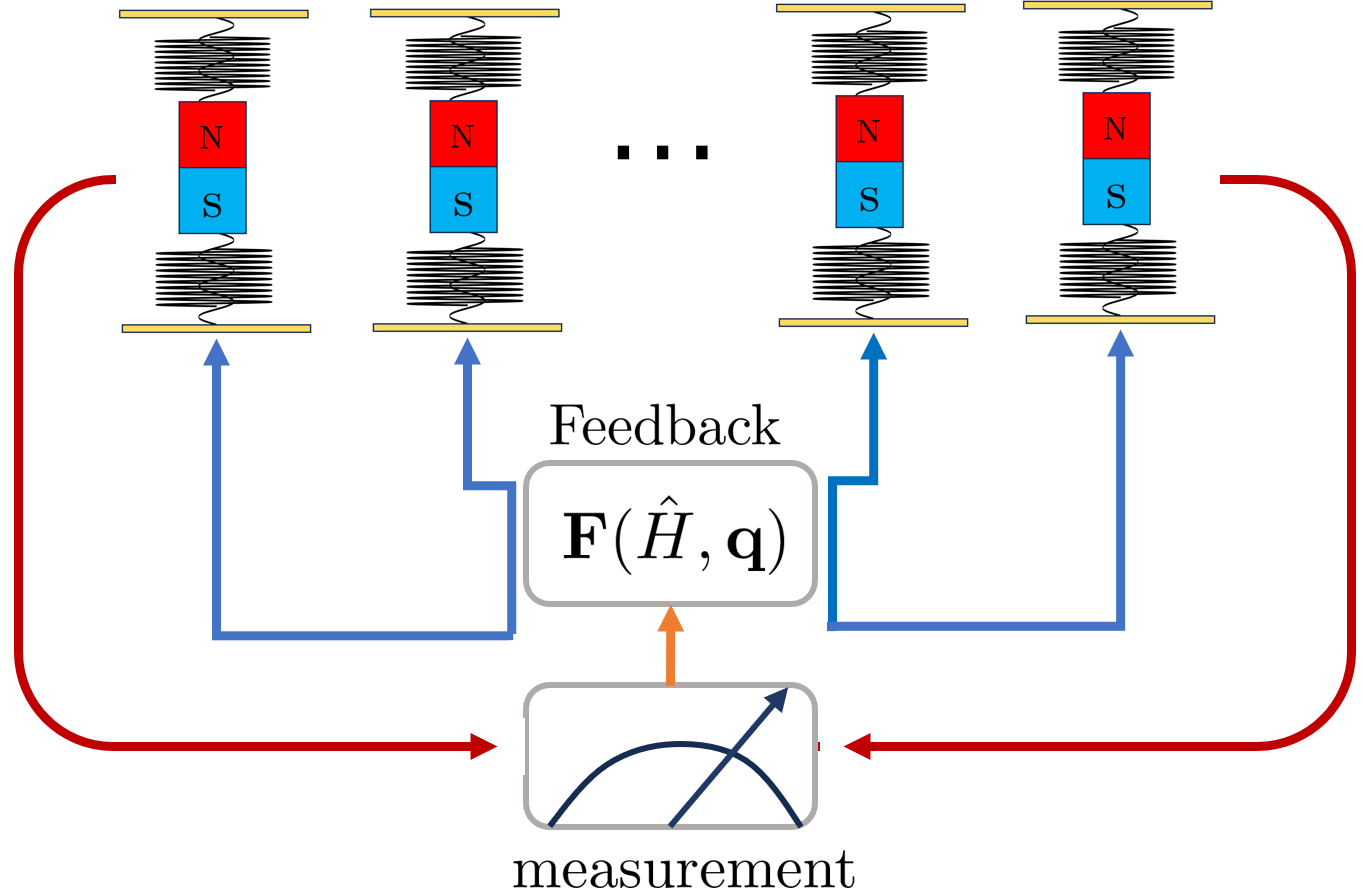}
	\caption{ Schematic diagram of a mechanical system to simulate the extended AA model. Real-time feedback forces $F$, which depend on the real-time position measurements $x_i$, are used to implement the target Hamiltonian $H$ by $F_n=-\partial H/\partial x_n$~\cite{anandwade2023-mechanical}.}
	\label{fig6}
\end{figure}

Specifically, the equations of motion for uncoupled and identical harmonic oscillators are
$
m \dot{x}_i(t)=p_i(t),\, \dot{p}_i(t)=-m \omega^2 x_i(t), 
$
where $x_i(t)$ and $p_i(t)$ are the position and momentum of the $i$-th oscillator, respectively, $\omega$ is the natural oscillation frequency, and $m$ is the mass. 
To couple these oscillators as a basis for simulating different Hamiltonians, we now add feedback forces to the system such that the equation of motion becomes
\begin{equation}
	\label{newton}
	\dot{X}_i=P_i, \quad \dot{P}_i=-\omega^2 X_i+f_i,
\end{equation}
where $f_i$ is a linear function of $(X_1,X_2,\cdots,X_n,P_1,P_2,\cdots,P_n)$ with $X_i\equiv-\omega^2 x_i(t)$ and $P_i\equiv \frac{-\omega^2}{m}p_i(t)$. The Newtonian equation (\ref{newton}) can be mapped to the Heisenberg equation by introducing the classical complex variables
\begin{equation}
	\dot{\alpha}_i=-i \omega \alpha_i+\frac{i}{\sqrt{2 \omega}} f_i
	=-i \omega \alpha_i+\frac{i}{\sqrt{2 \omega}} 
	\sum_j M_{ij}(\alpha_j+\alpha_j^*),
\end{equation}
where $\alpha_i \equiv \sqrt{\frac{\omega}{2}} X_i+i \sqrt{\frac{1}{2 \omega}} P_i$,
and the feedback term is expressed as a linear function of $f_i=\sum_j M_{ij}(\alpha_j+\alpha_j^*)$. 
This equation can be regarded as the Heisenberg equation of motion derived from a quantum mechanical Hamiltonian (with $\hbar=1$)
\begin{equation}
	\label{Hquantum}
	\hat{H}=\sum_i \omega \hat{\alpha}_i^{\dagger} \hat{\alpha}_i-\frac{1}{\sqrt{2 \omega}} \sum_{i, j}  \hat{\alpha}_i^{\dagger} \hat{M}_{i j} (\hat{\alpha}_j+\hat{\alpha}_j^{\dagger}),
\end{equation}
where $\hat{\alpha}_i^{\dagger}$ and $\hat{\alpha}_i$ are creation and annihilation operators for site $i$ obeying bosonic commutation relations, with $\alpha_i=\left\langle\hat{\alpha}_i\right\rangle$ and $\alpha_i^*=\langle\hat{\alpha}_i^{\dagger}\rangle$.
Assuming the feedback is sufficiently weak compared to the natural oscillations, the time-dependence of the complex amplitude is still given by $\alpha_i(t) \propto e^{-i \omega t}$. This allows us to apply the rotating-wave approximation (RWA) to simplify Eq.~(\ref{Hquantum}) as
\begin{equation}\label{Hrwa}
	\hat{H}_{\rm RWA}=\sum_i \omega \hat{\alpha}_i^{\dagger} \hat{\alpha}_i-\frac{1}{\sqrt{2 \omega}} \sum_{i, j}  \hat{\alpha}_i^{\dagger} \hat{M}_{i j} \hat{\alpha}_j.
\end{equation}
From Eq.~(\ref{Hrwa}), we can design a nonlocal hopping Hamiltonian by controlling $M_{ij}$, which depends on the feedback force and real-time position measurement.

\subsection{Topoelectrical circuit}
\begin{figure}[tb]
	\centering
	\includegraphics[width=1\linewidth]{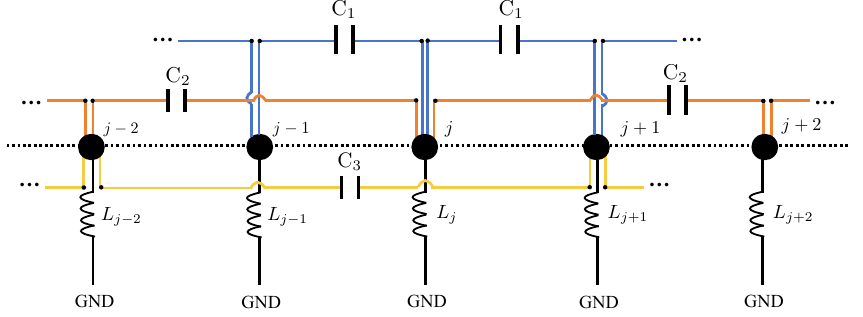}
	\caption{ Scheme for using a circuit to simulate the extended AA model. The capacitor $C_n$ is connected between two nodes with distance $n$, simulating the nonlocal hopping. The inductor $L_j$ connected to each node controls the on-site potential.}
	\label{fig7}
\end{figure}

We now turn to the implementation of the extended Aubry-Andr\'{e} model in topoelectrical circuits. The possibility of using circuits to simulate novel states of matter~\cite{albert2015a-circuit} has achieved tremendous success~\cite{Ningyuan2015,Lee2018,Helbig2020,wang2023} due to electronic components meeting almost all required characteristics. Crucially, their exceptional connection flexibility enables arbitrary-range couplings, making them ideal for nonlocal systems.

The general approach of circuit simulation is to implement Hamiltonian $H$ through the circuit Laplacian $\mathcal{J}$~\cite{Lee2018}, which determines the behavior of the circuit. Our theoretical design is shown in Fig.~\ref{fig7}. In that circuit, node $(j)$ is connected to node $(j\pm n)$ by a capacitor $C_n$ ($0<n\leq R$), and each node is grounded via an inductor $L_{j}$.
The relationship between the input current $I_{j}$ and voltage $V_{j}$ of the node is given by Kirchhoff's law,
\begin{equation}
	\label{circuit}
	I_{j} = i\omega \left[ \left(\sum_n C_{|n-j|}-\frac{1}{\omega^2L_{j}} \right)V_{j} - \sum_{l} C_{|j-l|} V_{l}\right] .
\end{equation}

Kirchhoff's law is expressed in matrix form as $I=\mathcal{J}V$, where $I$ and $V$ are vectors composed of the input currents and electrical potential of the circuit nodes, respectively. The off-diagonal elements $\mathcal{J}_{jl}=-i\omega C_{|j-l|}$ are determined by $C_{|j-l|}$ between nodes $(j)$ and $(l)$, while the diagonal elements $\mathcal{J}_{jj}=i\omega (\sum_n C_{|n-j|} -\frac{1}{\omega^2L_j})$ are controlled via inductance $L_{j}$. By selecting $C_{n}=C$ for $0<n\leq R$ and $\sum_n C_{|n-j|} -\frac{1}{\omega^2L_j} = -\frac{\lambda C}{J}\cos{(2\pi\beta j + \delta)}$, we can ensure that the Laplacian and Hamiltonian (\ref{Hamiltonian}) have the same mathematical form. 

In addition to being highly tunable, the circuit is also easy to measure. By measuring the voltage of the circuit nodes relative to ground, one can obtain the admittance bands~\cite{Helbig2019} (eigenvalue spectrum of the Laplacian $\mathcal{J}$). The inverse of the Laplacian, the Green's function $G=\mathcal{J}^{-1}$, can be understood as the response function of the circuit. When we inject current to a certain node, the current vector is represented as $I=[0,\ldots,I_m,\ldots,0]^{T}$. Since $V=GI$, we have $G_{nm}=V_n/I_m$. Therefore, by measuring the voltage, we can reconstruct the Green's function $G$, and then use numerical calculation to obtain the admittance spectrum and eigenstates.
}

\section{Conclusion}

We have explored particle transport in an extended Aubry-Andr\'e model with system-scale hopping, uncovering rich interplay between { localization, nonlocality, and gap-crossing spectral structure}. The quantum dynamics of particles in the insulating region exhibits flat-band-like localization in the absence of disorder, and further evolves into multi-point localization under a quasiperiodic potential. In contrast, particles in the metallic region remain diffusive but confined, giving rise to \rd{an MI coexistence phase}, where transport is allowed only within the metallic region and sharply suppressed in the insulating region. These domains are separated by \rd{sharp spatial boundaries}.
Meanwhile, system-scale hopping induces { gapped spectra with gap-traversing branches} in the AA model. Different from the conventional exponential localization, we uncovered unconventional { gap-crossing domain-wall modes} whose spatial probability distributions form comb-like patterns with multiple equidistant peaks. These { gap-crossing domain-wall modes} enable coherent, nonlocal, multi-point transport over long distances.
Our results demonstrate that system-scale hopping not only reshapes the character of localization and transport but also qualitatively modifies the spatial structure of { gap-crossing modes}. These findings establish extreme nonlocality as a new mechanism for exploring engineered coherent transport and { gap-crossing transport phenomena} in quasiperiodic systems, with direct experimental relevance to electrical circuits~\cite{Lu2018,wang2023,albert2015a-circuit}, photonic and mechanical metamaterials~\cite{Lustig2019,Tian2024,anandwade2023-mechanical,Wang2024}, and programmable quantum simulators~\cite{chen2025a-bryce,meier2018-bryce,Tian2025}. { A proper topological characterization of such gap-crossing modes in system-scale nonlocal hopping systems remains an interesting open question for future work.}

\begin{acknowledgments}
This work is supported by the National Natural Science Foundation of China (Grant No. 12125402, No. 92265208, No. 12074428, No. 12534016, and No. 12504585), the National Key R\&D Program of China (Grant No. 2022YFA1405301), and the Beijing Natural Science Foundation (Grant No. Z240007).  K.S. acknowledges the China Postdoctoral Science Foundation (Grant No. 2025M783357).
\end{acknowledgments}

\bibliography{citation}

\end{document}